\documentclass{PoS}
\usepackage{amsmath}
\usepackage{subcaption}
\usepackage[numbers,sort&compress]{natbib}

\DeclareMathAlphabet{\mathcal}{OMS}{cmsy}{m}{n}
\DeclareSymbolFontAlphabet{\mathnormal}{letters}
\DeclareSymbolFont{letters}{OML}{ztmcm}{m}{it}
\title{QCD Equation of State From a Chiral Hadronic Model Including
  Quark Degrees of Freedom}%
\ShortTitle{QCD EoS From a Chiral Hadronic Model Including Quarks}
\author{\speaker{Philip Rau}$^{1,2}$, Jan Steinheimer$^3$, Stefan
  Schramm$^1$, Horst St\"ocker$^{1,2,4}$\\%
  $^1$Frankfurt Institute for Advanced Studies (FIAS),
  Ruth-Moufang-Str.\ 1, 60438 Frankfurt am Main, Germany\\
  $^2$Institut f\"ur Theoretische Physik, Goethe - Universit\"at,
  Max-von-Laue Str.\ 1, 60438 Frankfurt am Main, Germany\\
  $^3$Lawrence Berkeley National Laboratory, Berkeley, CA 94720, USA\\
  $^4$GSI Helmholtzzentrum f\"ur Schwerionenforschung GmbH,
  Planckstr.\ 1,  64291 Darmstadt, Germany\\
  E-mail: \email{rau@th.physik.uni-frankfurt.de}}
\abstract{This work presents an effective model for strongly
  interacting matter and the QCD equation of state (EoS). The model
  includes both hadron and quark degrees of freedom and takes into
  account the transition of chiral symmetry restoration as well as the
  deconfinement phase transition. At low temperatures $T$ and baryonic
  densities $\rho_B$ a hadron resonance gas is described using a
  SU(3)-flavor sigma-omega model and a quark phase is introduced in
  analogy to PNJL models for higher $T$ and $\rho_B$. In this way, the
  correct asymptotic degrees of freedom are used in a wide range of
  $T$ and $\rho_B$. Here, results of this model concerning the chiral
  and deconfinement phase transitions and thermodynamic model
  properties are presented. Large hadron resonance multiplicities in
  the transition region emphasize the importance of heavy-mass
  resonance states in this region and their impact on the chiral
  transition behavior. The resulting phase diagram of QCD matter at
  small chemical potentials is in line with latest lattice QCD and
  thermal model results.}
\FullConference{8th International Workshop on Critical Point and Onset
  of Deconfinement\\March 11-15\\Napa, California, USA}
\begin{document}
\section{Introduction}
\label{sec:intro}
Detailed knowledge on the phase diagram of strongly interacting matter
is still scarce and the study of QCD matter, in particular at finite
temperature $T$ and baryon density $\rho_B$ or chemical potential
$\mu_B$ attains increasing interest. In addition to the well known
ground-state properties, from heavy-ion experiments with highest
collision energies, nuclear matter is known to exhibit properties of a
nearly perfect fluid at very high $T$. At $\mu_B = 0$, lattice QCD
predicts a smooth cross-over transition for the restoration of chiral
symmetry at $T \approx 160$~MeV. For all $\mu_B > 0$, lattice QCD
suffers from the fermion sign-problem which hinders to find solutions
in this region. To obtain information on QCD matter at finite
potentials, expansion methods can be used. However, until today
neither from experiment nor from theory there is a clear picture
whether the transition possibly changes to first order and a critical
end point exists. Therefore, the phase structure of QCD matter under
extreme conditions remains object of an ongoing and lively scientific
debate.\par
This work presents results on the QCD phase diagram from an effective
chiral flavor SU(3) model and contrasts them to recent results from
the lattice. The model includes all known hadronic degrees of freedom
as well as a quark-gluon phase implemented in a PNJL-like approach.
\section{Chiral Effective Model}
\label{sec:model}
The $SU(3)$-flavor chiral effective model~\cite{model} unifies a
$\sigma$-$\omega$-model for the hadronic phase using a non-linear
realization of chiral symmetry and a PNJL-like approach for the quark
phase. The model is based on a mean field Lagrangian $\mathcal{L} =
\mathcal{L}_{\rm kin} + \mathcal{L}_{\rm int} + \mathcal{L}_{\rm mes}$
which includes the particles' kinetic energy, the interaction of
baryons with scalar ($\sigma$, $\zeta$) and vector ($\omega$, $\phi$)
meson fields $\mathcal{L}_{\rm int} = -\sum_i \bar{\psi_i} [ \gamma_0
( g_{i\omega} \omega^0 + g_{i\phi} \phi^0 ) + m^*_i ] \psi_i$. Index
$i$ runs over the quark flavors $u$, $d$, $s$, and all known baryons
(octet, decuplet, and heavy-mass baryon resonances with $m \le
2.6$~GeV). The effective particle masses $m_{i}^* = g_{i\sigma}\sigma
+ g_{i\zeta}\zeta + \delta m_i$ and their effective chemical
potentials $\mu^*_i = \mu_i - g_{i \omega} \omega - g_{i \phi} \phi$
are generated dynamically (except for small explicit masses $\delta
m_i$) by the coupling to the scalar and the vector fields
respectively. With increasing $T$ and $\rho_B$, $\sigma$ decreases,
the effective masses decline, and chiral symmetry is restored. The
couplings of the baryon octet are fixed such as to reproduce
well-known vacuum masses and nuclear saturation properties.  All quark
couplings are chosen according to the additive quark model and such as
to restrain free quarks from the ground state.  The baryon resonance
couplings (including the decuplet) are scaled by the coefficients
$r_s$, $r_v$ to the respective couplings of the nucleons via $g_{B
  \sigma, \zeta } = r_{s} \cdot g_{N \sigma, \zeta}$ and $g_{B \omega,
  \phi} = r_{v} \cdot g_{N \omega, \phi}$. To obtain a smooth
cross-over at $\mu = 0$, the scalar coupling is chosen $r_s \approx
1$. The resonance vector coupling $r_v$ has a large impact on the
transition behavior and on the resulting phase diagram. For reasonable
values $r_v \approx 1$, the chiral transition is a smooth cross-over
in the whole $T$--$\mu$ plane.\par
The last term of the model Lagrangian includes the vector and scalar
meson self interactions as well as explicit chiral symmetry breaking
$\mathcal{L}_{\rm mes} = \mathcal{L}_{\rm vec} +\mathcal{L}_{0} +
\mathcal{L}_{\rm ESB}$. All thermodynamic quantities are derived from
the grand canonical potential $ \Omega / V = -\mathcal{L}_{\rm int} -
\mathcal{L}_{\rm mes} + \Omega_{\rm th} / V - U_{\rm Pol} $, including
the thermal contribution $\Omega_{\rm th}$ of hadrons and quarks which
couple to the Polyakov loop $\Phi$. Furthermore, $\Omega$ includes the
Polyakov loop potential $U_{\rm Pol}$ defining the quark dynamics in
the model~\cite{polpot}. The shift of degrees of freedom from a hadron
resonance gas (HRG) to a pure gas of quarks and gluons is realized by
excluded volume effects assuming finite-volume hadrons and point-like
quarks. Due to the excluded volume, at high $T$ and $\mu$ hadrons
vanish from the system and the QGP-phase prevails. Introducing volume
correction factors ensures thermodynamics consistency.
\begin{figure}[t]
  \centering 
  \begin{subfigure}[t]{.486\linewidth}\centering
    \includegraphics[width=1\columnwidth]{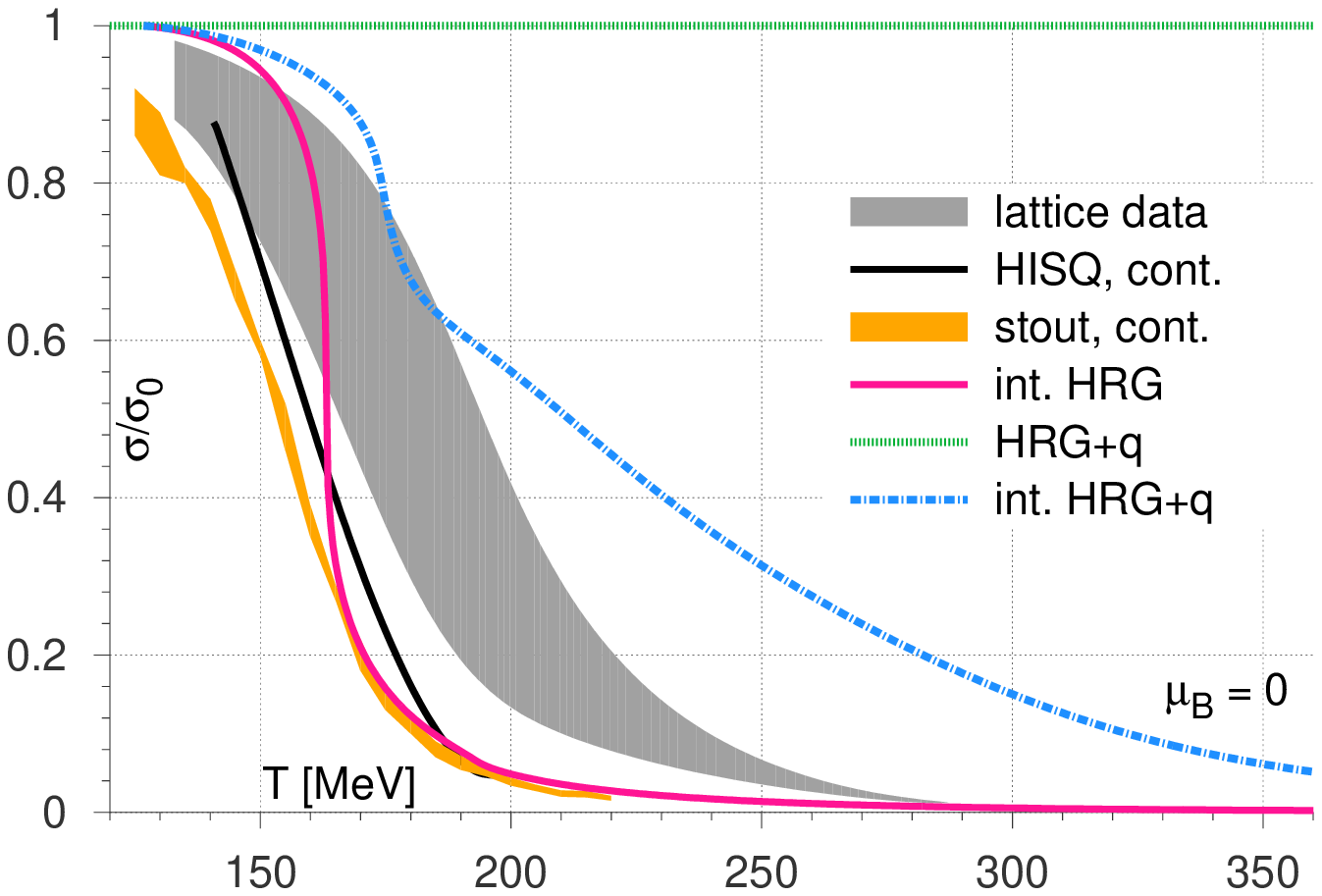}
    \caption{}\label{fig:mod_sigma_T}
  \end{subfigure}
  \hfill%
  \begin{subfigure}[t]{.486\linewidth}\centering
    \includegraphics[width=1\columnwidth]{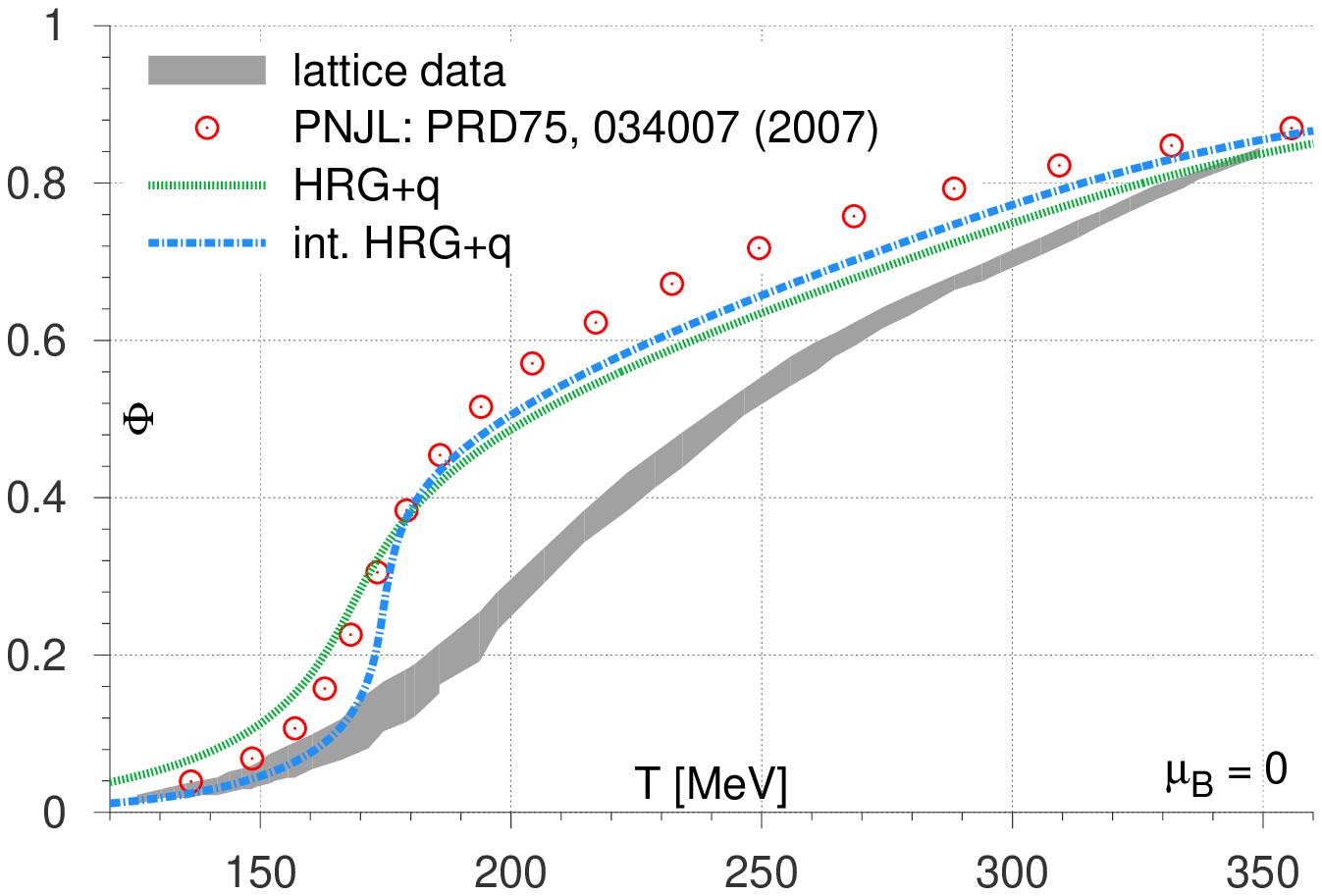}    
    \caption{}\label{fig:mod_pol_T}
  \end{subfigure}
  \caption{Normalized chiral condensate (I) compared to older (gray
    band) and newest continuum extrapolated lattice
    data~\cite{latdat}. The Polyakov loop from the effective model
    (analogously to PNJL models~\cite{polpot}) shows a more rapid
    shift to deconfined quarks as in lattice QCD (II).}
  \label{fig:fig01}
\end{figure}
\section{Results}
\label{sec:results}
Fig.~\ref{fig:fig01} (I) shows the chiral order parameter normalized
by its ground state value $\sigma/\sigma_0$ as a function of $T$. The
Polyakov loop order parameter $\Phi$ is shown in Fig.~\ref{fig:fig01}
(II). The chiral model results are compared to lattice QCD with
different fermion actions~\cite{latdat}. The illustration contrasts
different parameter sets and particle compositions of the chiral
model. A pure interacting HRG without quarks (int.\ HRG) exhibits the
steepest (and continuous) incline of $\sigma$ within a small
temperature range. In line with most recent lattice results, a
critical temperature $T_c = 164$~MeV is determined.\par
\begin{figure}[t]
  \centering 
  \begin{subfigure}[t]{.449\linewidth}\centering
    \includegraphics[width=1\columnwidth]{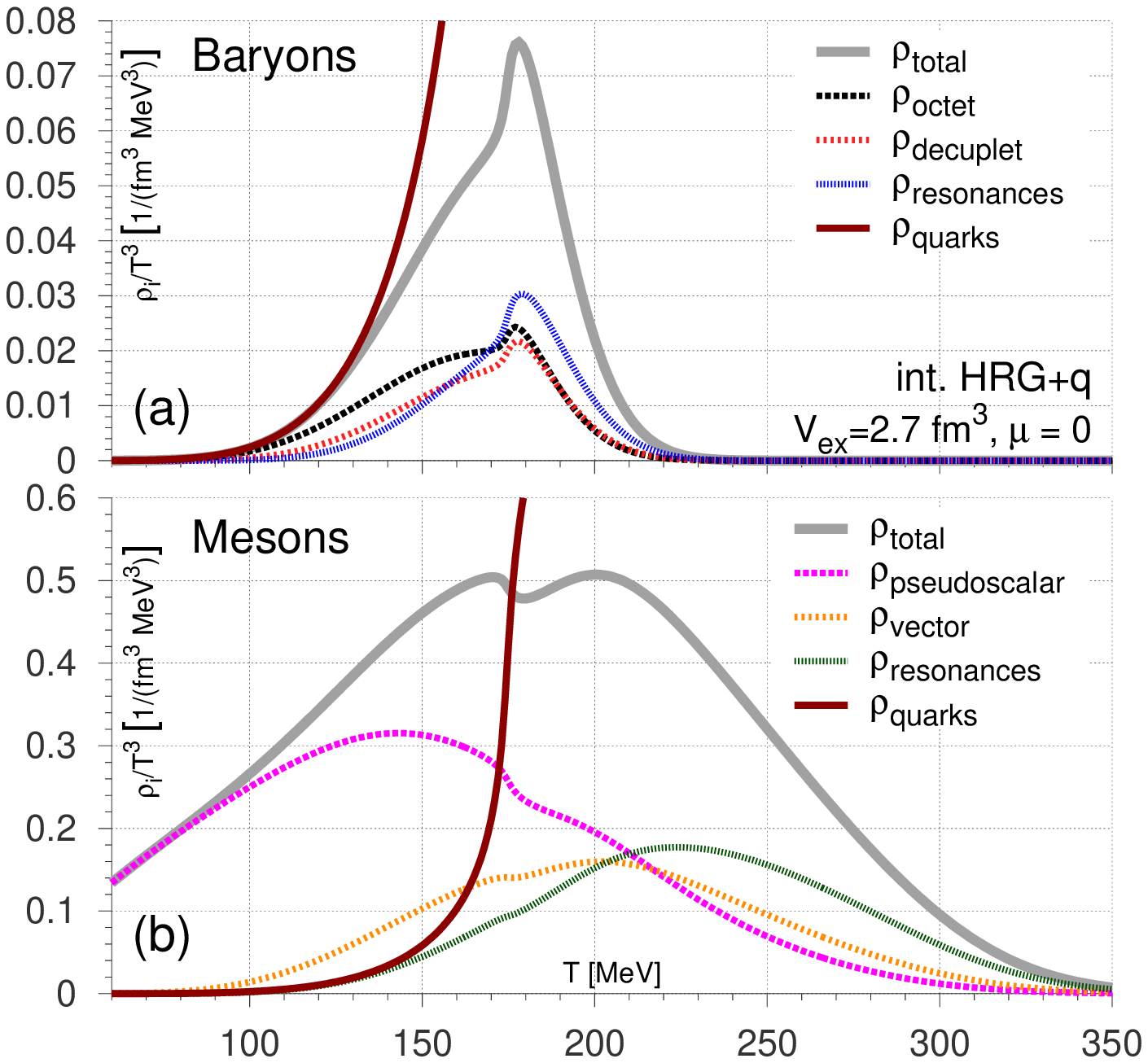}
    \caption{}\label{fig:spec_mult}
  \end{subfigure}
  \hfill%
  \begin{subfigure}[t]{.449\linewidth}\centering
    \includegraphics[width=1\columnwidth]{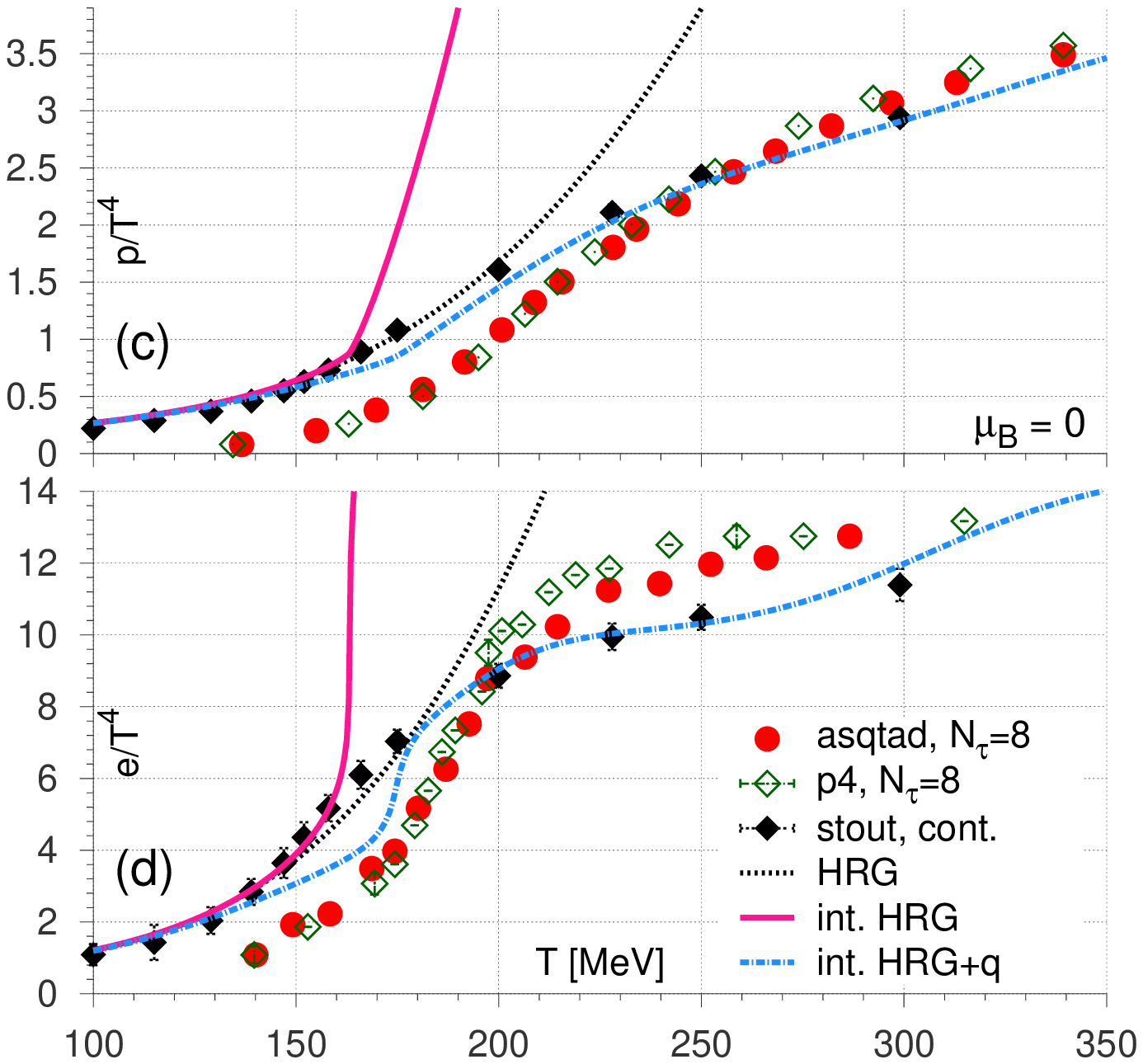}    
    \caption{}\label{fig:thermodyn}
  \end{subfigure}
  \caption{Total baryon number density of baryons (a), mesons (b), and
    quarks divided by $T^3$ at $\mu = 0$. Illustrated are the total
    densities (gray lines) and the density shares of different
    particle species. Panels (c) and (d) depict pressure and energy
    density over $T^4$ from different model scenarios contrasted to
    lattice QCD as functions of $T$.}
  \label{fig:fig02}
\end{figure}

Other model scenarios additionally include the PNJL-like quark phase
with either no interactions between meson fields and particles (HRG+q)
or full interactions (int.\ HRG+q). While in the non-interacting case
$\sigma$ stays at its ground state value, the decrease of $\sigma$ in
the int.\ HRG+q scenario is rather steep in the beginning. However, as
soon as quarks are abundant above the critical Polyakov temperature
$T_0 = 175$~MeV (Fig.~\ref{fig:fig01} (II)), the slope of $\sigma(T)$
flattens significantly. Comparing the full model results to pure HRG,
apparently, hadrons define the slope of $\sigma(T)$ at least up to
$T_c$ and, thus, the chiral transition is driven by hadrons to a large
extent. Due to quarks lacking an eigenvolume, they are preferentially
populated at $T \ge T_c$ and the number of quarks exceeds the number
of hadrons (Fig.~\ref{fig:fig02} (a)). In this region, the
significantly smaller quark scalar coupling causes $\sigma$ to
decrease much slower. However, as illustrated in Fig.~\ref{fig:fig01}
(II), the rapid increase of quark multiplicities at $T_c$ generates a
distinctly faster increase of $\Phi$ than predicted by lattice QCD
(compiled in the gray band). A fundamental discrepancy between $\Phi$
from PNJL models~\cite{polpot} and lattice QCD~\cite{latdat} in the
transition region is apparent.\par
Quantifying the impact of heavy-mass resonances at the phase
transition, Fig.~\ref{fig:fig02} (I) shows the baryon number density
of baryons (a) and mesons (b) together with quarks divided by $T^3$ at
$\mu = 0$. The total densities (gray line) are broken down into
contributions from low- and intermediate-mass particles as well as
heavy-mass resonances and the red line illustrates the total quark
density.\par
Below $T_c$, low-mass baryons from the octet, mainly nucleons,
represent the largest share of the total baryon density
(Fig.~\ref{fig:fig02} (a)). This changes as soon as $m^*$ of the
resonances decline due to a decreasing $\sigma$-field. At $T \ge T_c$,
heavy-mass resonances are the most abundant baryons in the
system. Even though $\Delta$-resonances reach rather large
multiplicities with increasing temperature, the combined density
contribution of the decuplet stays lowest at all $T$.  Regarding the
mesons (Fig.~\ref{fig:fig02} (b)), due to the abundant light-mass
pions below $T_c$, pseudoscalar meson multiplicities are dominant in
this region. At $T_c$, the fast rising quark number and the
pseudoscalar meson mass scaling $m_{\rm mes}^{*2} \sim 1/\sigma$ cause
pseudoscalar meson multiplicities to decrease as well as a perceivable
dip in the total meson density. Higher temperatures are accompanied by
an increase in the vector meson densities and a distinctive appearance
of heavier meson resonances which become the most abundant mesons at
$T \approx 210$~MeV. This finding of large resonance multiplicities at
$T_c$ and slightly above underlines the substantial impact of
heavy-mass resonances on the chiral transition. The effect of
resonances is substantial even in the presence of quarks, where
excluded volume effects suppress hadrons at high $T$, and should be
even larger in the pure HRG. Due to their large influence, heavy-mass
hadron resonances must be considered when studying the phase
transition region. Notwithstanding this study only considers $\mu =
0$, this conclusion should also hold true in the whole
$T$--$\mu_B$-plane.\par
Fig.~\ref{fig:fig02} (II) shows the pressure $p$ (c) and the energy
density $e$ (d) divided by $T^4$. For both quantities the interacting
pure HRG perfectly reproduces continuum extrapolated lattice QCD
results up to $T_c$. Also the non-interacting HRG yields a
qualitatively good agreement up to slightly higher temperatures. In
the whole temperature range, the fully interacting model including
quarks yields reasonable results with only minor deviations from
lattice in $e/T^4$ around $T_c$. This dip in $e/T^4$ at $T = (155 \pm
15)$~MeV is caused by the hard-core repulsion of the hadrons. The
quark density is just starting to rise sharply and can not compensate
the lower hadron contribution in this region. Therefore, the energy
density flattens up to $T_c$ and rises sharply again thereafter.\par
\begin{figure}[t]
  \centering 
    \includegraphics[width=.683\columnwidth]{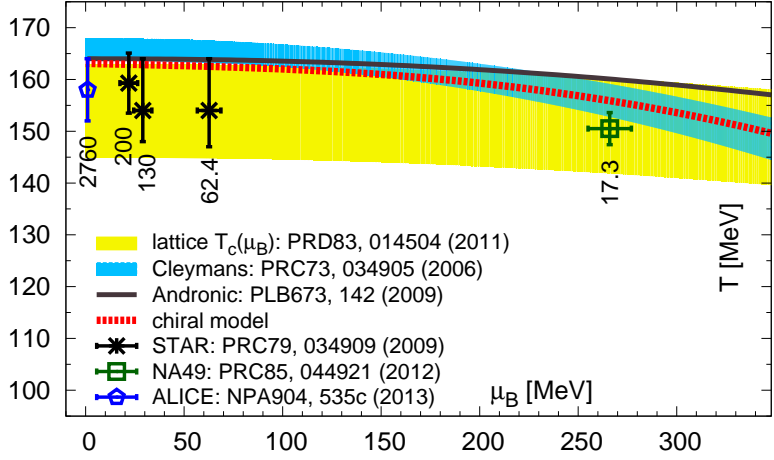}
    \caption{Chiral transition from lattice QCD~\cite{phasediag}
      (yellow band) and from the chiral effective model (red line) at
      small $\mu_B$. The estimates for the transition are compared to
      freeze-out curves from statistical models and to data from
      thermal model fits~\cite{FO-curve} for SPS to LHC energies
      ($\sqrt{s_{\rm NN}}$ in GeV).}
    \label{fig:fig03}
\end{figure}
Due to the large number of hadron degrees of freedom, the chiral
transition is a smooth cross-over even at large $\mu_B$. In comparison
to lattice QCD extrapolations~\cite{phasediag} and thermal model fits
of experimental results~\cite{FO-curve}, the phase transition line
from the chiral model (Fig.~\ref{fig:fig03}) yields a plausible
estimate located at the upper boundary of lattice QCD results.\par
In summary, this work presented a unified effective model which not
only provides the correct ground state properties but also describes
the transition from a confined HRG to a deconfined quark-gluon phase
reasonably well and in line with latest lattice QCD results. An
extracted hadron-quark EoS is to be used within dynamic models to
study nuclear matter properties in heavy-ion collisions.


\begin{thebibliography}{99}
\bibitem{model}%
  J.~Boguta and H.~St\"ocker, Phys.\ Lett.\ {\bf B120}, 289 (1983); %
  P.~Papazoglou {\it et al.}, Phys.\ Rev.\ {\bf C57}, 2576 (1998); %
  Phys.\ Rev.\ {\bf C59}, 411 (1999); %
  J.~Steinheimer {\it et al.}, arXiv:0909.4421; %
  P.~Rau {\it et al.}, arXiv:1302.3836.%
%
\bibitem{polpot}%
  C.~Ratti {\it et al.}, Eur.\ Phys.\ J.\ {\bf C49}, 213 (2007); %
  S.~Roessner {\it et al.}, Phys.\ Rev.\ {\bf D75}, 034007 (2007).%
%
\bibitem{latdat}%
  A.~Bazavov {\it et al.}, Phys.\ Rev.\ {\bf D80}, 014504 (2009); %
  Y.~Aoki {\it et al.} JHEP {\bf 0906}, 088 (2009); %
  M.~Cheng {\it et al.}, Phys.\ Rev.\ {\bf D81}, 054504 (2010); %
  S.~Borsanyi {\it et al.}, JHEP {\bf 1011}, 077 (2010); %
  A.~Bazavov {\it et al.}, Phys.\ Rev.\ {\bf D85}, 054503 (2012).%
%
\bibitem{phasediag}%
  O.~Kaczmarek {\it et al.}, Phys.\ Rev.\ {\bf D83}, 014504 (2011); %
  A.~Bazavov {\it et al.}, Phys.\ Rev.\ {\bf D85}, 054503 (2012).%
%
\bibitem{FO-curve}%
  J.~Cleymans {\it et al.}, Phys.\ Rev.\ {\bf C73}, 034905 (2006); %
  D.~Zschiesche {\it et al.}, J.\ Phys.\ {\bf G34}, 1665 (2007); %
  A.~Andronic {\it et al.}, Phys.\ Lett.\ {\bf B673}, 142 (2009); %
%
  B.~I.~Abelev {\it et al.}  [STAR], Phys.\ Rev.\ {\bf C79}, 034909
  (2009); %
  F.~Becattini {\it et al.}, Phys.\ Rev.\ {\bf C85}, 044921 (2012); %
  A.~Andronic {\it et al.}, Nucl.\ Phys.\ {\bf A904-905}, 535c
  (2013).%
\end{thebibliography}
\end{document}